\documentclass[%
prl%
 ,twocolumn%
 ,superscriptaddress%
,tightenlines%
,amssymb,amsmath,nobibnotes, aps, prl]{revtex4}
\usepackage{dcolumn}
\usepackage{bm}
\usepackage[dvips]{graphicx}
\begin{document}
\title{NMR Evidence for Spin Canting in a Bilayer $\nu = 2$ Quantum Hall System}
\author{N. Kumada}
\author{K. Muraki}
\affiliation{NTT\,Basic\,Research\,Laboratories,\,NTT\,Corporation,\,3-1\,Morinosato-Wakamiya,\,Atsugi\,243-0198,\,Japan}
\author{Y. Hirayama}
\affiliation{NTT\,Basic\,Research\,Laboratories,\,NTT\,Corporation,\,3-1\,Morinosato-Wakamiya,\,Atsugi\,243-0198,\,Japan}
\affiliation{SORST-JST, 4-1-8 Honmachi, Kawaguchi, Saitama 331-0012, Japan}
\affiliation{Department of Physics, Tohoku University, Sendai 980-8578, Japan}

\date{Version: \today}

\begin{abstract}
We investigate the electron spin states in the bilayer quantum Hall system at total Landau level filling factor $\nu =2$ exploiting current-pumped and resistively detected NMR.
The measured Knight shift, $K_S$, of $^{75}$As nuclei reveals continuous variation of the out-of-plane electronic spin polarization between nearly full and zero as a function of density imbalance.
Nuclear spin relaxation measurements indicate a concurrent development of an in-plane spin component.
These results provide direct information on the spin configuration in this system and comprise strong evidence for the spin canting suggested by previous experiments.
\end{abstract}
\maketitle


Bilayer quantum Hall (QH) systems have found rich electronic phases that emerge as a consequence of the interplay between different degrees of freedom.
In particular, at total Landau level filling factor $\nu =2$, where two of the four lowest Landau levels split by the Zeeman and interlayer tunnel couplings are occupied, the competing spin and layer degrees of freedom lead to rich magnetic phases.
When the interlayer coupling is weak, the system behaves like two independent single-layer $\nu =1$ systems, resulting in a ferromagnetic (F) ground state with spins in each layer aligned parallel to the magnetic field by intralayer interactions and the weak Zeeman coupling.
When the tunneling is strong, on the other hand, the interlayer antiferromagnetic coupling leads to a spin-singlet (SS) ground state resembling the single-layer $\nu =2$ state.
Inelastic light scattering \cite{Pellegrini,Pellegriniscience} has shown a mode softening indicating the existence of another state between these two states.
Theory \cite{Zheng,DasSarma1997,DasSarma1998,Demler,Breybias,MacDonald1999} suggested a canted antiferromagnetic (CAF) state, where spins in the two layers have antiparallel in-plane components and parallel out-of-plane components.
Transport \cite{SawadaPRL,Geer,Fukuda} and capacitance \cite{Khrapai} measurements have also shown a phase transition.
Recently, a nuclear spin relaxation measurement \cite{Kumadascience} has revealed a gapless spin excitation mode (Goldstone mode) indicative of the in-plane antiferromagnetic order \cite{DasSarma1997,DasSarma1998}.
Although these experiments are consistent with the CAF state, direct information on the spin configuration, such as the spin polarization $P_z$, has not been reported.

Nuclear magnetic resonance (NMR) is a powerful probe for $P_z$ because the hyperfine interaction between electron and nuclear spins shifts the nuclear resonant frequency by the Knight shift $K_S$, which is proportional to $P_z$.
In two-dimensional electron systems (2DESs), a weak signal resulting from a small number of nuclei in contact with the 2DES restricts standard NMR measurements to multiple-quantum-well systems \cite{Barrett,Tycko,Kuzma,Melinte}.
Recently, a $K_S$ measurement on a single-layer system has been achieved \cite{OStern}, where the NMR signal from nuclei in a quantum well (QW) is resistively detected.
However, the method cannot be used to measure $K_S$ in a well-developed QH state, where the longitudinal resistance $R_{xx}$ is vanishingly small regardless of the nuclear spin polarization.

In this Letter, we report the measurement of $K_S$ in the bilayer $\nu =2$ QH state using current-pumped and resistively detected NMR.
To measure $K_S$ in well-developed QH states, we exploit a new technique based on selective radio frequency (rf) irradiation synchronized with the switching between two electronic states; that is, NMR is performed on the bilayer $\nu =2$ QH state while nuclear spin polarization is induced and then resistively detected in the single-layer $\nu =2/3$ state.
The measured $K_S$ of $^{75}$As nuclei reveals that $P_z$, that is, the out-of-plane spin polarization, changes continuously between nearly full at the balanced density condition and zero at large density imbalance.
The $K_S$ data, combined with the nuclear spin relaxation rate, show that the intermediate values of $P_z$ are accompanied by the development of the in-plane antiferromagnetic order.
These results present strong evidence for spin canting that intervenes between the F and SS states.

The sample used in this study consists of two 200-\AA-wide GaAs QWs separated by a 31-\AA-thick Al$_{0.3}$Ga$_{0.7}$As barrier.
The sample has a low temperature mobility of 3.1$\times 10^{6}$\,cm$^2/$V\,s and a tunneling energy gap of $\Delta _{\rm SAS}=8$\,K at a total density of $2.7\times 10^{11}$\,cm$^{-2}$.
The filling factors in the front and back layers ($\nu _{\rm f}$ and $\nu _{\rm b}$) can be tuned independently using the front- and back-gate biases.
We tune the bilayer $\nu =2$ system through different magnetic phases by changing the density imbalance, $\delta \equiv (\nu _{\rm f}-\nu _{\rm b})/(\nu _{\rm f}+\nu _{\rm b})$, between the layers at fixed magnetic field $B$ \cite{Breybias,MacDonald1999} rather than by changing $B$ at $\delta =0$ as originally proposed \cite{Zheng,DasSarma1997}.
The sample was mounted in a dilution refrigerator with a base temperature of 50\,mK and subjected to a constant magnetic field of $B=5.5$\,T perpendicular to the 2DES.
To apply an rf field perpendicular to the static $B$-field, a wire loop was placed around the sample, which resulted in a higher temperature of 120\,mK for the $K_S$ measurement.

\begin{figure}[t]
\begin{center}
\includegraphics[width=0.95\linewidth]{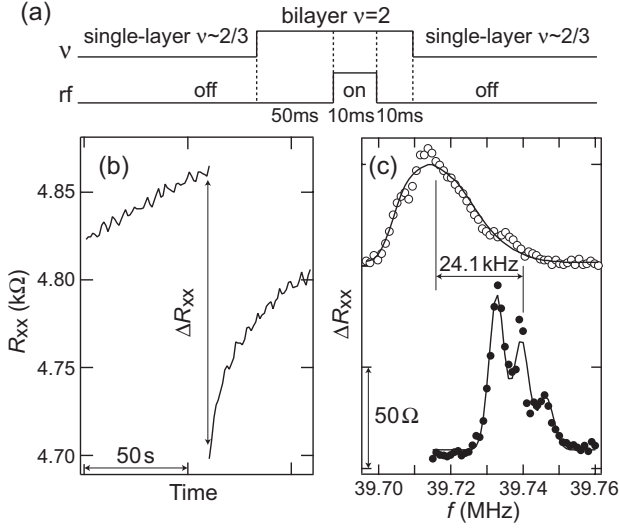}
\caption{
(a) Timing diagram of the current-pumped and resistively detected NMR.
(b) Time evolution of $R_{xx}$ at $(\nu _{\rm f}, \nu _{\rm b})=(\sim 2/3,0)$, illustrating its response to rf irradiation during the short (70 ms) excursion to the bilayer $\nu =2$ state.
(c) NMR spectra of $^{75}$As at depletion (solid circles) and for the $\nu =2$ state at $\delta =0.02$ (open circles).
Lines are fits, from which $K_S=24.1$\,KHz is obtained.
Traces are vertically offset for clarity.
}
\label{sequence}
\end{center}
\end{figure}

\begin{figure}[t]
\begin{center}
\includegraphics[width=0.98\linewidth]{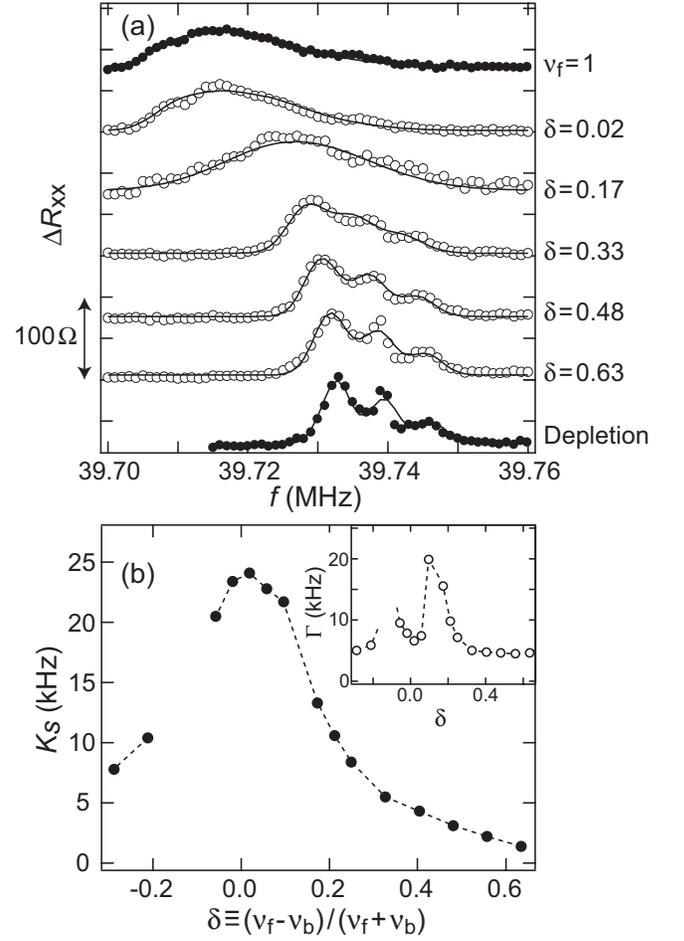}
\caption{
(a) NMR spectra of $^{75}$As for several values of density imbalance $\delta $.
For comparison, spectra at depletion (bottom) and for the single-layer $\nu =1$ state (top) are shown by closed circles.
Solid lines represent the results of fitting (see text for details).
Traces are vertically offset for clarity.
(b) $K_S$ determined from the fitting plotted as a function of $\delta $.
Inset shows $\Gamma $ vs $\delta $.
}
\label{Ks_nu2}
\end{center}
\end{figure}

Figure\,\ref{sequence} describes our NMR technique.
First, the nuclear spin polarization is current-induced in the front layer by setting the electronic system to the single-layer $\nu =2/3$ state, $(\nu _{\rm f}, \nu _{\rm b})=(\sim 2/3,0)$, which is manifested by $R_{xx}$ increasing slowly over a typical time constant of 10 min (not shown here) \cite{Kronmuller1999,HashimotoPRB}.
After $R_{xx}$ has saturated, the electronic system is tuned to the bilayer $\nu =2$ state and maintained for a short period of time (70\,ms), during which the rf field with a frequency $f$ is applied for 10\,ms [Fig.\,\ref{sequence}(a)].
Then the electronic system is switched back to $(\nu _{\rm f}, \nu _{\rm b})=(\sim 2/3,0)$, where we measure the change in $R_{xx}$ of the single-layer $\nu =2/3$ state that has occurred after the short excursion to the bilayer $\nu =2$ state [Fig.\,\ref{sequence}(b)] \cite{lockin}.
As shown in Fig.\,\ref{sequence}(b), $R_{xx}$ decreases after the rf irradiation, and we define $\Delta R_{xx} (>0)$ as the size of the resistance drop.
By plotting $\Delta R_{xx}$ as a function of $f$, an NMR spectrum is obtained [Fig.\,\ref{sequence}(c)] \cite{relaxation}.

Open circles in Fig.\,\ref{sequence}(c) shows an example of NMR spectra of $^{75}$As nuclei, taken for the bilayer $\nu =2$ state close to the balanced density condition ($\delta =0.02$).
The spectrum is asymmetrically broadened, similar to that for the $\nu =1/3$ state obtained by standard NMR in a multiple-quantum-well system \cite{Kuzma}.
As detailed below, this reflects the spatial variation of the local electron density (and hence of the Knight shift) in the direction normal to the QW plane.

To determine the magnitude of the Knight shift, we measured the NMR spectrum in the absence of the 2DES.
This was done by completely depleting the 2DES (instead of tuning it to the bilayer $\nu =2$ state) during the rf irradiation.
The result is shown in Fig.\,\ref{sequence}(c) by closed circles.
The spectrum is free from the Knight shift and so appears at higher frequencies.
It is narrower and, as a result, the threefold splitting due to the quadrupole moment of $^{75}$As nuclei is resolved.
By fitting this spectrum using a Gaussian function $g(\Gamma _0;x)$ with a full width at half maximum of $\Gamma _0=4.1$\,kHz, we obtain the center-peak frequency $f_c=39.740$\,MHz, the quadrupole splitting $\Delta f=6.8$\,kHz, and the relative amplitude of the three peaks, $(A_{-1}, A_0, A_1)=(1.4, 1.0, 0.49)$.
We used these values to fit the NMR spectra in the presence of the 2DES following the method described in Ref.\,\cite{Kuzma}, where the effects of the spatially varying density due to finite QW thickness are taken into account.
The fitting function is
\begin{eqnarray}
\nonumber I(f)& = &\sum _{m=-1}^{1}A_m\int_{f_c}^{f_c+K_S}dx\sqrt{\frac{x-f_c}{f_c+K_S-x}}\\
\nonumber & &\times g(\Gamma ;f-x-m\Delta f),
\label{fit}
\end{eqnarray}
where $K_S$ and $\Gamma $ are used as adjustable parameters.
As shown by the solid line in Fig.\,\ref{sequence}(c), the fit is reasonably good, yielding $K_S=24.1$\,kHz and $\Gamma =6.6$\,kHz \cite{error}.

\begin{figure}[t]
\begin{center}
\includegraphics[width=0.85\linewidth]{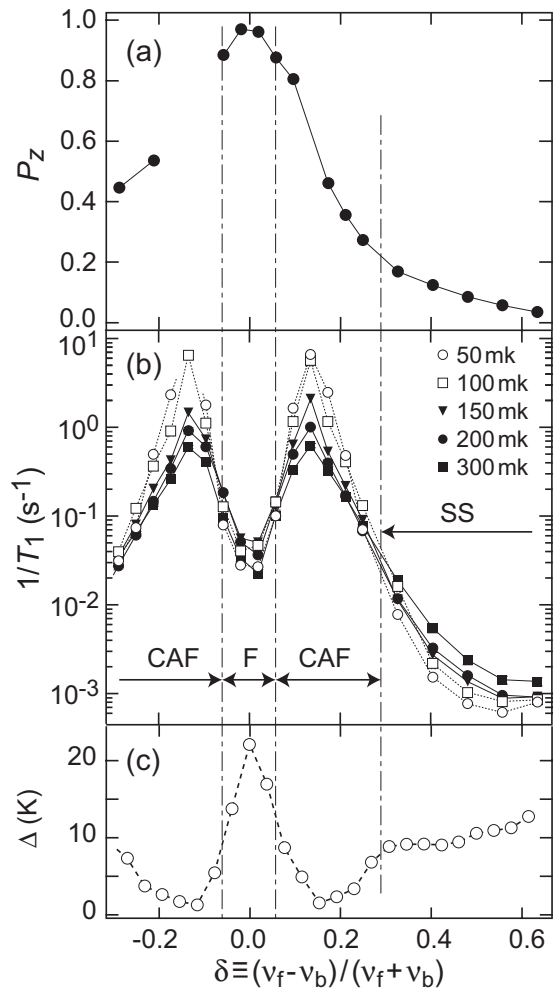}
\caption{
(a) Electron spin polarization $P_z$, (b) nuclear spin relaxation rate $1/T_1$ for several temperatures and (c) charge excitation gap $\Delta $ as a function of $\delta $.
Vertical dashed-dotted lines represent the phase boundaries determined from the temperature dependence of $1/T_1$.
}
\label{KsT1delta}
\end{center}
\end{figure}

NMR spectra for different magnetic phases were obtained by performing similar measurements for different values of $\delta $.
The results are shown in Fig.\,\ref{Ks_nu2}(a) together with the spectrum at depletion.
As a reference for a fully polarized state, the NMR spectrum for the single-layer $\nu =1$ state, $(\nu _{\rm f}, \nu _{\rm b})=(1,0)$, is also shown.
As the data demonstrate, at small $|\delta |$ the spectrum for the bilayer $\nu =2$ system is very similar in both position and shape to that for the single-layer $\nu =1$ state, while at large $|\delta |$ it much resembles that at depletion.
These results confirm that the bilayer $\nu =2$ system is fully-polarized at small $|\delta |$ and unpolarized at large $|\delta |$, consistent with theory \cite{Breybias,MacDonald1999}.
Between these states the spectra transform continuously.
In Fig.\,\ref{Ks_nu2}(b), we plot $K_S$ determined from the fitting as a function of $\delta $.
$K_S$ varies continuously between the values at full and null polarizations.
As we show later, transport measurements show a well-developed QH effect at low temperatures over the whole range of $\delta $, with the activation energy remaining finite.
We emphasize that simple pictures, such as that for the integer QH effect or the composite-fermion picture for the fractional QH effect, suggest that each QH state should have a definite spin polarization given by the number of occupied spin-up and -down levels \cite{Kukushkin}.
We also point out that the observed NMR spectra cannot be explained by the phase separation into domains of ferromagnetic and symmetric states, which would result in spectra consisting of two independent contributions from fully-polarized and unpolarized regions \cite{OStern}.
The observed continuous evolution of the spin polarization with the QH effect maintained, in turn, suggests a nontrivial QH state that allows for continuous transformation of its spin configuration.

To convert $K_S$ into $P_z$, we need to take into account the fact that not only the spin polarization but also the electron density in each layer changes with $\delta $.
Since in our experimental scheme the NMR is resistively detected using a single-layer $\nu =2/3$ state formed in the front layer, the measured $K_S$ reflects the polarization and the density in the front layer.
Therefore, $K_S$ for a fully polarized system is expected to vary with $\delta $ in the following way:
\begin{eqnarray}
\nonumber \overline{K_S}(\delta )=\nu _{\rm f}K_S^{(\nu =1)}=(1+\delta )K_S^{(\nu =1)},
\label{convert}
\end{eqnarray}
where $K_S^{(\nu =1)}=24.7$\,kHz is the measured $K_S$ for the single-layer $\nu =1$ state.
The spin polarization is given by $P_z(\delta )=K_S(\delta )/\overline{K_S}(\delta )$, which is plotted in Fig.\,\ref{KsT1delta}(a) as a function of $\delta $ \cite{asymmetry}.

We now compare $P_z$ with the nuclear spin relaxation rate $1/T_1$  [Fig.\,\ref{KsT1delta}(b)] and the charge excitation gap $\Delta $ [Fig.\,\ref{KsT1delta}(c)] of the same sample measured in a different cool down.
$1/T_1$ was measured using a current-pump and resistive-detection technique \cite{Kumadascience,KumadaPRL2} and $\Delta $ was determined from the temperature dependence of $R_{xx}$, i.e., $R_{xx}\propto \exp (-\Delta /2T)$, where $T$ is the temperature.
The minimum value of $\Delta $ is 1.6\,K, which results in the QH effect being preserved at the temperature for the $K_S$ measurement (120\,mK).
As shown in our previous study \cite{Kumadascience}, the temperature dependence of $1/T_1$ can be used to determine the phase boundaries between different magnetic phases (vertical lines in Fig.\,\ref{KsT1delta}).
We note that $1/T_1$ reflects the zero-frequency fluctuations of the {\it in-plane} component of electron spins.
Thus, $1/T_1$ and $P_z$ provide complementary information on the spin state, the former and the latter reflecting the in-plane and out-of-plane components of the local magnetization, respectively.
The regions for $|\delta |<0.06$ and $|\delta |>0.3$, where $P_z$ is nearly full and zero, are identified as the F and SS phases, respectively.
For $0.06<|\delta |<0.3$, $1/T_1$ is enhanced and diverges with decreasing temperature, indicating the development of the in-plane antiferromagnetic order parameter and its Goldstone mode.
This coincides with the region where $P_z$ varies over a wide range from $P_z=0.9$ to 0.2.
The intermediate values of $P_z$ accompanied by the development of the in-plane spin order attests to the canted spin configuration.
The continuous change in $P_z$ represents the evolution of the spin canting as follows.
When the CAF phase is approached from the F phase, electron spins, which are initially aligned normal to the plane, start to cant.
As a result, $P_z$ decreases and, at the same time, the in-plane spin component starts to develop.
Upon further increasing $\delta $, as a result of increased spatial overlap of the different spin states, the in-plane spin component starts to decrease at some point and eventually vanishes together with $P_z$, where the system enters the SS phase.

Although the data are broadly consistent with theory \cite{Zheng,DasSarma1997,DasSarma1998,Demler,Breybias,MacDonald1999}, there are some unexpected features.
First, $P_z$ is not constant at $P_z=1.0$ in the F phase, dropping to $P_z\approx 0.9$ at its phase boundary.
We note that at $\delta \neq 0$ the filling factor in each layer deviates from unity as $(\nu _{\rm f},\nu _{\rm b})=(1+\delta ,1-\delta )$.
This will create skyrmions and antiskyrmions in opposite layers, charged spin textures in the single-layer $\nu =1$ state constituting the $\nu =2$ bilayer F state, thereby reducing $P_z$ \cite{Barrett}.
We speculate that the binding of skyrmions and antiskyrmions preserves the QH effect as we observed.
These spin-textured quasiparticles, which are expected to be present even at $\delta =0$ due to disorder, are responsible for $1/T_1$ in the F phase \cite{Kumadascience,KumadaPRL2}.
On the other hand, in the SS phase, $P_z$ is not completely zero.
As the data in Fig.\,\ref{KsT1delta}(b) show, in this phase the energy gap for spin fluctuations decreases toward zero as the phase boundary is approached.
At finite temperatures, this not only increases $1/T_1$ but will also lead to finite $P_z$.
The sudden drop of $\Delta $ at $\delta =0.3$, combined with the fact that this occurs exactly where $1/T_1$ changes its temperature dependence, lends further strength to the existence of a phase boundary despite the continuous variation of $P_z$.
Rather, continuous variations of both $P_z$ and $1/T_1$ across the phase boundaries indicate that the transitions are of second order, consistent with theory \cite{Zheng,DasSarma1997,DasSarma1998,Demler,Breybias,MacDonald1999}.
The relevance of the low-energy spin excitations probed here via $1/T_1$ to the continuous phase transitions revealed by $P_z$ presents an interesting challenge for theory.
Another interesting observation is that $\Delta $ shows a shoulder at the CAF-SS boundary and a minimum in the middle of the CAF phase, while Hartree-Fock theories \cite{Breybias,MacDonald1999}, instead, predict only weak cusps at the boundaries.
We speculate that spin-texture excitations not incorporated in the existing theories may explain these discrepancies.

Finally, we briefly discuss $\Gamma $, which is mainly due to spatial inhomogeneity of the in-plane electron density.
As shown in the inset of Fig.\,\ref{Ks_nu2}(b), $\Gamma $ is enhanced in the CAF state.
Since $P_z$ in the CAF state sensitively depends on intralayer and interlayer interactions, inhomogeneity of system parameters, such as the electron density, the layer distance, or $\Delta _{\rm SAS}$, introduce inhomogeneous $P_z$, leading to broad NMR spectra.
Detailed investigation of the spectra would provide spatial information, which may reveal a theoretically predicted spin Bose-glass phase \cite{Demler}.

\begin{acknowledgments}
The authors are grateful to T. Saku for growing the heterostructures, and T. Ota and T. Fujisawa for fruitful discussions.
\end{acknowledgments}


\begin{thebibliography}{99}
\bibitem{lockin}
To circumvent the finite response time of the lock-in amplifier, we employed the $R_{xx}$ value at a few seconds after returning to the $\nu =2/3$ state. As the data in Fig.\,1(b) show, this hardly affects the results.
\bibitem{relaxation}
Although longitudinal relaxation of nuclei during the short excursion causes a constant offset to the NMR spectrum, it does not affect its frequency dependence.
\bibitem{error}
The accuracy of the fitting is limited by the approximate form of the fitting function, where the slight asymmetry of the QW wave function is neglected.
The resultant error in $K_S$ is less than 10\%, which does not affect our conclusion.
\bibitem{asymmetry}
$P_z$ is slightly asymmetric about $\delta =0$, which may be due to an asymmetric sample structure: the modulation doping is carried out only on the front layer and back-layer electrons are induced by a positive back gate bias.
\end{thebibliography}
\end{document}